
\input harvmac
\def\figflag{I}
\global\newcount\figno \global\figno=1
\newwrite\ffile
\def\pfig#1#2{Fig.~\the\figno\pnfig#1{#2}}
\def\pnfig#1#2{\xdef#1{Fig. \the\figno}%
\ifnum\figno=1\immediate\openout\ffile=figs.tmp\fi%
\immediate\write\ffile{\noexpand\item{\noexpand#1\ }#2}%
\global\advance\figno by1}

\def\tfig#1{Fig.~\the\figno\xdef#1{Fig.~\the\figno}\global\advance\figno by1}

\def\figI{I}
\newdimen\tempszb \newdimen\tempszc \newdimen\tempszd
\newdimen\tempsze
\ifx\figflag\figI
\input epsf
\def\epsfsize#1#2{\expandafter\epsfxsize{
 \tempszb=#1 \tempszd=#2 \tempsze=\epsfxsize
     \tempszc=\tempszb \divide\tempszc\tempszd
     \tempsze=\epsfysize \multiply\tempsze\tempszc
     \multiply\tempszc\tempszd \advance\tempszb-\tempszc
     \tempszc=\epsfysize
     \loop \advance\tempszb\tempszb \divide\tempszc 2
     \ifnum\tempszc>0
        \ifnum\tempszb<\tempszd\else
           \advance\tempszb-\tempszd \advance\tempsze\tempszc \fi
     \repeat
\ifnum\tempsze>\hsize\global\epsfxsize=\hsize\global\epsfysize=0pt\else\fi}}
\epsfverbosetrue
\fi
\def\ifigure#1#2#3#4{
\midinsert
\vbox to #4truein{\ifx\figflag\figI
\vfil\centerline{\epsfysize=#4truein\epsfbox{#3.eps}}\fi}
\narrower\narrower\noindent{\bf #1:} #2
\endinsert
}

\Title{\vbox{\hbox{HUTP--93/A028}}}{\vbox{\centerline{{Strings and
Singularities}}}}
\vskip .2in

\centerline{Cumrun
Vafa\foot{Based on talk presented at Salamfest, March 1993,
Trieste.}}
\vskip .2in \centerline{Lyman Laboratory of Physics, Harvard University}
\centerline{Cambridge, MA 02138, USA}
\vskip .3in
We review aspects of spacetime singularities from the view
point of string theory.  Examples considered include cosmological,
cosmic string and black-hole singularities.  We also discuss
the consistency of viewing
black-holes as excited states of fundamental strings.
\Date{10/93}

It is a great pleasure to speak in this conference in honor of Professor
Abdus Salam.  I have a deep respect for him, not only
because of his important contributions
to theoretical physics but also for his complete
devotion to the cause of
advancing science in third world countries.

In this talk I will discuss aspects of space time singularities
in the context of string theory.

One of the main hints about the incompleteness of
classical general theory of
relativity comes from the fact that almost all interesting solutions
have physical singularities.  Even if we start with a non-singular
universe typically in the past or in the future there are times
when we hit a singularity and do not know how to continue physics
beyond.  Let us consider three examples of singularities:

{\bf{A-Cosmological}}

In the standard big bang cosmology, the universe is believed to have
started in a situation of infinite curvature--
this is also the reason that the time before big bang
cannot be defined.  Moreover at
the singularity the temperature is infinite which is rather
unphysical.

{\bf{B-Cosmic strings}}

There are other kinds of examples of curvature singularities
which are expected to appear in general relativity, the simplest
example of which is that encountered for an infinitely thin
cosmic string. In this case the geometry of space
is singular only along the string itself, where
the geometry is conical.  This means that the transverse
geometry looks flat
except some piece of the angular region is missing.

{\bf{C-Astrophysical Singularities:  Black holes}}

In most astrophysical objects which are compact enough we encounter
singularities.  The most well known example is that of a black-hole
singularity.  These are very important to understand because
star evolution could naturally lead to black holes, and
thus their physical existence
is an inescapable consequence of relativity combined with the
the existence of compact objects.

One of the standard reactions to the existence of such singularities
is to express a hope that quantum effects should naturally take
care of these infinities.  After all, quantum phenomena smooth things
out, and so one should expect that smearing these infinities will
be a quantum resolution to the paradox.  Before the advent of string
theories such hopes were expressed without a realistic hope of testing
it, as we did not have any viable candidates of a quantum theory of gravity.
With the emergence of string theory as essentially the only viable
candidate of a quantum theory of gravity, we are thus forced to face
the question:  How does quantum strings resolve the singularities
inherent in the general theory of relativity?

It turns out that string theory does have ways to cure these
singularities at least in the context of simple examples.
Though the complete story has not yet emerged we consider these
examples as typical of how string deals with singularities.
  The way string theory
resolves these singularities is not just by quantum smearing out but
more surprisingly even {\it classically} strings do not see a singularity.
We will now discuss these in turn in the context of concrete examples.

\subsec{Strings and Cosmology}

String cosmology has been modeled by putting strings in
a box with periodic boundary conditions \ref\bv{R. Brandenberger
and C. Vafa, Nucl. Phys. B316 (1988) 28.}.
Consider a box of size $L$ in all directions, measured in string
(=planck) units (\tfig\FigOne).  Consider some matter in the box
in an equilibrium state at temperature
$T$.  Assume that as universe evolves the temperature changes
in an adiabatic way (i.e. to maintain constant entropy). We thus
have a function $T(L)$.  Then in point particle theory as $L\rightarrow
0$, the temperature diverges $T\sim 1/L$.  Moreover the dynamics
of relativity implies that (in a flat space with zero cosmological constant)
$L\sim t^{1/2}$ for small time $t>0$ after the big bang.  The curvature
of space time blows up as $t\rightarrow 0$.

\ifigure\FigOne{The strings in a box can
be in different types of states:  They could be wound
around the box (W) or oscillating inside the box, or
moving from one side of the box to another (P), or some
combination thereof.}{Fig1}{2.0}

The story is very different in the context of string theory:  There
are two major modifications:  1-There is a very important symmetry
for a string in a box: Strings in a box of size $L$ and in a box of
size $1/L$ behave exactly the same
\ref\duali{K. Kikkawa and M. Yamasaki, Phys. Lett. B149 (1984) 357\semi
N. Sakai and I. Senda, Prog. Theor. Phys. 75 (1986) 692\semi
V. Nair, A. Shapere, A. Strominger and F. Wilczek, Nucl. Phys. B287 (1987)
402\semi B. Sathiapalan, Phys. Rev. Lett. 58 (1987) 1597\semi
E. Silverstein, {\it Large-Small
Equivalence in String Theory},
hepth-9201009\semi
G. Moore, {\it Finite in All Directions}, hepth-9305139.}\bv!
 There is no experiment we can
perform do distinguish in which geometry the string is living in.
This is amazingly strange.  The point is that string states are sensitive
to the size of the box $L$ only in two ways:  Their center of mass momentum
is quantized in inverse units of $L$, i.e.
$$P={n\over L}$$
This is also true for point particles.  However strings can also be wrapped
around the box.  Their energy is proportional to how many times
$m$ they wind around the box
$$W=m L$$
Note that if we exchange $L\leftrightarrow 1/L$, and if we relabel the
momentum and winding states $P\leftrightarrow W$, (i.e. $n\leftrightarrow m$)
we get exactly the same string spectrum.  Not only the spectrum
is the same but also all the interaction between states in
the box of size $L$ is the same as between the dual states in the box
of size $1/L$.  This example is the simplest example of a
phenomenon known as mirror symmetry, in which a string
propagating on two inequivalent manifolds behaves the same.  In
string theory such pairs of manifolds are thus
indistinguishable.
The duality symmetry in the context of the strings in a box
is very stringy
as it clearly depends on the existence of winding modes which
are absent in point particle theories.  It is also not a symmetry
of $p$-brane theories as can be easily seen by looking at
their spectra (for example for 2 dimensional extended objects
we get states with energy $1/L, L$ and $L^2$ and the duality
symmetry is thus lost).  So if we insist on the existence of the
duality symmetry $L\rightarrow 1/L$ we are forced to consider
one dimensional extended objects, i.e. strings.

We thus see that in the context of string theory
$L\rightarrow 0$ is no more singular than $L\rightarrow \infty$,
which is known to be non-singular.
The fact that there is this duality symmetry is nicely complemented
by the appearance of a maximum temperature for stings, known as the
Hagedorn temperature. Its existence is a deep fact of string theory and
is dictated by the exponential growth in the number of string states
with mass $M$, i.e.
$$d(M)\sim {\rm exp}(\beta_H M)$$
This gives rise to the limiting temperature
known as the Hagedorn temperature $T_H=\beta_H^{-1}$.

The dynamical
implementation of this
duality in string theory requires the existence of a scalar
field which is known as the dilaton
which transforms non-trivialy under $L\rightarrow 1/L$.
  Its existence
drastically
modifies the dynamics of general relativity, as we will now briefly
recall
in the context of string cosmology \ref\tv{
A.A. Tseytlin and C. Vafa, Nucl. Phys. B372 (1992) 443.}.
The existence of the dilaton field restores some of the intuition
one has with conservation of energy.  For example, in the context
of a flat universe with zero cosmological constant, having a constant
energy density of matter results in inflation, which looks
somewhat like the perpetual generation of energy.  One's naive
intuition would be that if there is a certain energy density,
expansion of the universe while keeping the energy density constant
should require `work' and thus there should be an obstacle to it.
The presence of dilaton restores this intuition in string theory!
  This suggests
in particular that inflation is not a natural phenomenon in string
theory (at least in the very early universe where dilaton field is not
frozen).

In the context of string cosmology, if we plot the total energy $E$
in the box as a function
of ${\rm log}L$ (see \tfig\FigTwo),

\ifigure\FigTwo{The total energy $E$ in the universe as a function of
the (logarithmic) size of the universe.  The state of the universe
is analogous to that of a particle rolling on the potential $E$ (with
some damping and modulation of the potential).  The fact that $E$ is
an even function of ${\rm log}L$ is a manifestation of scale duality.
  Note that the flat
region near L=1 is the Hagedorn phase of string, and seems
not to display any particularly singular behaviour.}{Fig2}{.95}

then the dependence of $L(t)$ can be inferred
roughly by thinking of the
state of the universe as a particle rolling on the potential $E$
(with some damping and modulation of the potential--see \tv ).
Note that, as mentioned above there is no singularity
as $L\rightarrow 0$.  Moreover the
Planckian region $L\sim 1$ (${\rm log}L\rightarrow
0$) is also a smooth region where the universe slowly rolls
on the almost flat potential $E$, where the flatness is a consequence
of the fact that in this region we are at the Hagedorn phase and
the energy is rather insensitive to $L$.
Also, it was using this model of string cosmology that
it was suggested in \bv\ that long strings
(or winding strings) if not decayed may be an obstacle
for the expansion of the universe
because $E$ will then grow linearly with $L$ in such cases and
this may be a natural
basis for the selection of 4 macroscopic dimensions for spacetime (two long
strings have a hard time finding each other in more
than 4 spacetime dimensions because 2+2=4)\foot{The dual picture
of this is well known:  Namely if we take $L\rightarrow 0$ then
the momentum modes now play the role of massive modes.
The momentum modes are just the excitation states for
point particle theory.  It
is well known that for point particle field theories with more
than 4 dimensions the theory is trivial (again the argument
is based on $2+2=4$) and so thermal equilibrium is difficult
to maintain.}.

\subsec{Cosmic Strings--Orbifold Singularities}
Let us now move to the singularities that appear in the case of
infinitely thin cosmic strings.  Consider having a cosmic string
with a deficit angle of $4\pi /3$ (see \tfig\FigThree).

\ifigure\FigThree{The transverse plane to a cosmic string
with deficit angle $4\pi /3$.  Note that we can describe
this geometry by identifying points on $R^2$ which differ
by $2\pi/3$ rotation.  In string theory we can get new
states, twisted states,
 whose two endpoints differ by the $Z_3$ action.}{Fig3}{1.5}

This means
that the angular variable goes from $0$ to $2\pi /3$.  It
is more convenient to imagine the presence of the full $2\pi$ angular
region but identify the three pieces each subtending a $2\pi /3$
angular region by dividing by a $Z_3$ symmetry, by identifying
$z\rightarrow exp(2\pi i/3)z $ where the plane perpendicular
to the direction of cosmic string is denoted by the complex parameter z.
This way of viewing the cosmic string is also known as orbifolds, i.e.,
the two dimensional space of the cosmic string can be viewed as a
quotient
$$R^2/Z_3$$
If one considers a field theory in this background one finds that unitarity
is lost at the quantum level.  The easiest way to see this is that
one can start with the Hilbert space of particles propagating on $R^2$
and project onto the $Z_3$ invariant subspace.  However, this cannot
possibly be consistent with unitarity because the
$Z_3$ non-invariant states that we have projected out will be
pair produced and so if we delete them from the spectrum we
just get a non-unitary theory.  If we include them we would be
back to $R^2$ and not the cosmic string configuration.

Strings have a beautiful way of resolving this inconsistency
\ref\orb{L. Dixon, J. Harvey, C. Vafa and E. Witten, Nucl. Phys.
B274 (1986) 285.}.  In the context of strings projecting
to the $Z_3$ invariant subspace of the Hilbert space is not
the end of the story:  We end up introducing two new sectors
replacing the states we projected out.  The reason for the presence
of these sectors is that we can have string states which are
closed up to $Z_3$ action (as depicted in Fig. 3).  These
states are closed strings when we recall that the points on the
plane differing by $Z_3$ action are identified with each other.
The two new sectors we get in this way are called the twisted
sectors.  The way string theory resolves the inconsistency
of field theory with unitarity is that in the loop amplitudes
instead of producing states which are not $Z_3$ invariant
and have been projected out, we produce pairs of twisted
string states.  This turns out to restore the unitarity.  Again
we see it is the extended nature of strings that allows
to have the extra sectors which restores unitarity, just as the
extended nature of winding modes in the previous example was
responsible for the duality invariance of strings.

\subsec{Black-holes and Strings}
Finally we would like to discuss how string theory comes to terms
with black holes.  This area needs to be further developed
and better understood.  However there is a two dimensional toy
model for black holes which is very similar to the actual
four dimensional theory \ref\witb{E. Witten, Phys. Rev. D44 (1991) 314.}\
which
can be discussed in the context of string theory.  Actually the
object of direct interest in string theory turns out to be
a supersymmetric version of the black-hole \ref\mv{S. Mukhi and C.
Vafa, {\it Two Dimensional Black-hole as a Topological Coset
Model of c=1 String Theory}, 1993 preprint, HUTP-93/A002 ,
TIFR/TH/93-01.}
(supersymmetry
here comes from the ghost degrees of freedom and is not part
of the matter theory).  The metric has the same singularity
structure as the 4d black hole;  it is given by
$$g={d\gamma d\bar \gamma \over 1-\gamma \bar \gamma}$$
(with a dilaton field turned on).  This back ground (with
the fermionic degrees of freedom included) can be described
as string propagation on the (supersymmetric) coset
$$SL(2)/U(1)$$
(with an appropriate `twisting' \mv ), where $\gamma $ and $
\bar \gamma$ can be viewed as particular (Gaussian) choice
of coordinates on $SL(2)$.  This theory is clearly
singular at $\gamma \bar \gamma =1$.  However as shown in
\mv\ this theory is equivalent to strings propagating on a circle
of radius 1, which is not singular!  In other words {\it the mirror
theory to strings on black hole is strings on circle}.  There is a way
to see this \ref\vv{C. Vafa and E. Verlinde, unpublished notes.}\
in the same explicit manner as in the duality for string in a box
using the techniques in \ref\ms{E. Martinec and S. Shatashvili
Nucl. Phys. B368 (1992) 338.}.  It is found that
by going to the conjugate variables to the $\gamma$'s, this can be
accomplished.  Let us briefly discuss how this comes about.
The world sheet action for string propagating on $SL(2)/U(1)$
(ignoring the fermionic contributions for simplicity)
can be written as
\eqn\act{\int \beta \overline
\partial \gamma +c.c. + \beta \overline \beta (1- \gamma \overline \gamma)}
Upon integrating $\beta ,\overline \beta $ out we get the usual sigma model
action for the black hole with the metric written above.
\eqn\bh{S=\int -{\overline  \partial \gamma
 \partial  \overline \gamma
\over (1-\gamma \overline \gamma)}}
However  we can
also integrate out the $\gamma$'s and we end up with
\eqn\ebh{\int {\overline  \partial \beta
 \partial  \overline \beta
\over \beta \overline \beta}+\beta \overline \beta }
Now rewriting $\beta \rightarrow \beta_0 exp(iX+\phi)$, with $\beta_0$
being a constant,
leads to the standard action for string propagating on the circle
(parametrized by $X$), where $\phi $ plays the role of the scale
of the worldsheet metric (Liouville field), and $\beta_0 \bar \beta_0
=\mu $
is identified as the cosmological constant on the worldsheet\foot{
It has been discovered \vv\ that the $\gamma$--description is related
to a ground ring generator, whereas $\beta $ is related to the tachyon
vertex.  So the two descriptions have complementary virtues.  Note also
that the tachyon states, which are given by $\beta^k \bar \beta $ and
$\beta \bar \beta^k $ in the $\beta$--description are represented in the
$\gamma $--description by $(\partial \bar \gamma)^k \bar \partial \gamma/(
1-\gamma \bar \gamma)^{k+1}$
and $\partial \bar \gamma (\bar \partial \gamma )^k/(
1-\gamma \bar \gamma)^{k+1}$
 which look like deforming the black hole geometry
by higher spin gravity theories--i.e. W-gravities.  In this connection
it is natural to expect that what corresponds in the
$\beta$ picture to tachyon condensates corresponds in the
the black-hole picture to turning on the higher spin
 W-gravity \vv .  This suggests that self-dual geometries
on the cotangent of the black-hole
may be the natural setup to describe black-hole deformations.}.
Note that the way string resolved this singularity
was {\it not} by a change of coordinate. In other words
$\beta $ cannot be written as a function of $\gamma $
alone. If this were the case we would simply have a change
of coordinates in the same sense as in the classical
theory of relativity.  Rather, what we have here is
$\beta (\gamma ,\partial \gamma )$, and so this change
of fields, does not translate to a geometrical operation
on the black-hole background.  Again we see that the sigma-model
description is responsible for avoiding any singularity (a
deeper understanding of this can be achieved by including the
fermionic terms \vv\ which mix with $\gamma ,\beta$ fields in the action).

We thus see that the black-hole configuration is not singular:
The $\gamma$-picture which looks singular is equivalent
to its `mirror' the $\beta$-picture, which is not singular.  Having
argued for the absence of physical singularity
in this 2 dimensional model for black-hole it is natural
to ask other questions relevant for $d$--dimensional black-holes.
There is a lot more to be learned from string theory about black-holes.
  One such question is in regard to its
thermal properties.
Consider a highly excited string state (a string state
with large oscillation number).  Such a
 state has a mass much bigger than its
size in Planck units (identifying string scale with Planck scale).
Therefore it is a candidate for a black-hole.  This is in line
with 't Hooft's idea of viewing fundamental elementary states
themselves as black-holes.  In this connection one would be naturally
led to ask questions relevant for black-hole dynamics from the
view point of strings.  One natural question is how
long does it take for a black-hole to evaporate?  This is thus
translated to the question of how long it takes for the highly excited
string state to decay.  What is the spectrum of massless states
one is left with?  Does the decay product of
a massive string state resemble a thermal distribution of massless
string modes?  If so how does the temperature of this distribution
depend on the mass of the excited string state?

At first sight this correspondence may raise some paradoxes.
 The first question is that of entropy of a black-hole:
It is natural to
 assume that the reason for the non-vanishing entropy of the black-hole
is that there is a lack of information as to how the energy $M$
is shared among various string states.  However this runs to immediate
problem:
The entropy of the black-hole is dimension
dependent and it goes as $exp(M^{d-2})$ for a $d$-dimensional
spacetime, whereas the Hagedorn behaviour of strings gives
$exp(M)$ for the entropy of a string state with mass $M$.  However due
to a very interesting observation of Susskind
\ref\sus{L. Susskind,{\it Some Speculations about Black Hole Entropy in String
Theory}, 1993 preprint,
hepth-9309145.}\
this is not necessarily a contradiction as there may be a mass renormalization
due to gravitational dressing, which
effectively replaces $M\rightarrow M^{d-2}$.
This may remove a fundamental obstacle
in identifying a massive string state with a black-hole.

 There are other interesting connections:
  If we compactify superstrings from
10 dimensions down to 4 on a six dimensional torus, and consider for a
fixed point in space, a string state which winds around one of the cycles
of the internal torus of length $L$.  Let us further assume that $L>>1$.
Then viewed from the four dimensional view point we have a state whose
mass is $M=L$ and whose extent is much less than $L$, and so is a black hole.
But we know that in string theory such a
winding state is absolutely stable (if it has no oscillations on it--otherwise
it would decay to a winding state with no oscillations).  Thus the question
is why such a black-hole does not decay.  The answer is simply
that it also carries a $U(1)$ charge equal to $Q=L$.  So we have
a state with $Q=M$, which is the condition for extremality of a
black-hole.  Such a black-hole is not expected to radiate
from four dimensional view point either.  In fact the situation is
more interesting, because string theory also predicts
that for any string state $M\geq Q$ (which in the conformal language
follows from the unitarity bound
$h\geq Q^2/2$), which is precisely the bound
expected for charged black-holes!  This lends further
support to the natural idea that string states themselves
may be viewed as black-holes\foot{The similar bound in the case
of heterotic strings is $M\geq \sqrt{Q^2-1}$, which shows that for
large enough $Q$ we get the usual bound expected on classical grounds,
but also there is no contradiction with the existence of massless
charged states in string theory.}.

These points suggest that we should identify all
the extremal black hole states with absolutely stable string
states.  We are thus naturally led to ask the question:  how
does one classify stable states in a string theory?
Moreover, if we bring two such states together, do we get
another stable state?  These questions are being currently
considered \ref\iv{K. Intriligator and
C. Vafa, work in progress.}.  The results are in line
with the intuition that if something is stable, there is some
kind of `charge' keeping it from absolute decay (i.e. the
state is the minimum energy state in a given charge sector).

\subsec{Other Examples of Singularity}
We have shown that many interesting examples of singularities
are harmless in string theory.  One should be careful
before making a sweeping generalization that all singularities
are harmless.  In fact there are examples of singular
backgrounds (manifolds) which
seem harmful in string theory.  These manifolds can be reached
with finite action from smooth ones.  An important
example of this happens for the quintic threefold at a
particular point on moduli $(\psi =1)$ (see \ref\quin{
P. Candelas, X.De le Ossa, P. Green and L. Parkes, {\it Essays on Mirror
Manifolds},
ed. by S. T. Yau,  International Press, 1992. }).
It would be very important to understand what happens to string
theory in such a limit, and whether it signals an incompleteness
of string theory.

I would like to thank K. Intriligator, L. Susskind and E. Verlinde
for valuable discussions.  This work was supported in part by
NSF grants PHY-87-14654 and PHY-89-57162 and a Packard fellowship.
\listrefs
\end